\documentclass[prb,showpacs]{revtex4}
\usepackage{amsfonts}
\usepackage{amssymb}
\usepackage{graphicx}
\usepackage{dcolumn}
\usepackage{amsmath}

\begin{document}

\title{Variational Approach for the Effects of Periodic Modulations on the
Spectrum of Massless Dirac Fermions}
\author{B. S. Kandemir and A. Mogulkoc}
\affiliation{Department of Physics, Faculty of Sciences, Ankara University, 06100\\
Tando\u{g}an, Ankara, Turkey}
\date{\today}

\begin{abstract}
In the variational framework, we study the electronic energy spectrum of
massless Dirac fermions of graphene subjected to one-dimensional oscillating
magnetic and electrostatic fields centered around a constant uniform static
magnetic field. We analyze the influence of the lateral periodic modulations
in one direction, created by these oscillating electric and magnetic fields,
on Dirac like Landau levels depending on amplitudes and periods of the field
modulations. We compare our theoretical results with those found within the
framework of non-degenerate perturbation theory. We found that the technique
presented here yields energies lower than that obtained by the perturbation
calculation, and thus gives more stable solutions for the electronic
spectrum of massless Dirac fermion subjected to a magnetic field
perpendicular to graphene layer under the influence of additional periodic
potentials.
\end{abstract}

\maketitle

Since the discovery of graphene by Novoselov et. al\cite{novoselov}, there
have appeared many theoretical studies on graphene and graphene based
nanostructures such as graphene quantum dots and graphene nanoribbons. There
are also few reviews which have been devoted to these structures\cite%
{rev1,rev2}. In most of these works, theoretical studies on the electronic
energy spectrum of these structures have focused, due to their linear
dispersion relation near $K$ points in the Brillouin zone\cite{wallace}, on
the continuum version of tight binding Hamiltonian, i.e., $2+1$ Dirac-Weyl
equation\cite{semenoff}, which has been confirmed by the experiments.\cite%
{zhang}

The theoretical considerations of Weiss oscillations in an
electrically modulated graphene were discussed by Peeters and
Matulis\cite{peeters}. They showed that, within the framework of
non-degenerate first order perturbation theory , these
non-relativistic oscillations are more pronounced in graphene as
compared with those found for its $2D$ non-relativistic
counterpart\cite{matulis}. Very recently, studies on these
oscillations in a magnetically modulated graphene have also been
reported\cite{tahir,peeters2,Quan,Chiu,Luca}.

On the one hand, from the non-relativistic point of view, after Weiss et al.%
\cite{weiss}, theoretical studies on magnetoresistance
oscillations in two-dimensional electron gas structures subjected
to a periodic potential have been intensively studied for a long
time within the frame work of perturbation theory\cite{pert}. It
is shown that these oscillations have as a common origin the
oscillating bandwidth of the modulation-broadened Landau
bands\cite{gerhardts}. On the other hand, role of one-dimensional
periodic potentials is crucial itself in graphene. Very recently,
it was shown that, by a Muffin-Tin type\cite{park} one-dimensional
periodic potential, velocity of charge carriers can be controlled
in graphene. These type potentials allow one to create various
types of graphene superlattices and to control the flow of charge
carriers in an non-invasive way.

In this paper, we present a variational analysis of the combined
effects of electric and magnetic potentials on the electronic
energy spectrum of a Dirac-Weyl like electron in graphene. Our
method is based upon the use of solutions of massless Dirac
fermions in an uniform magnetic field as trial wave functions in
the presence of external potentials. Analytical results obtained
in this paper indicate that the variational method is more
efficient than the other previously used methods, i.e.,
non-degenerate first-order perturbation theory.

\begin{figure}[p]
\includegraphics*[ height=14cm,width=8cm]{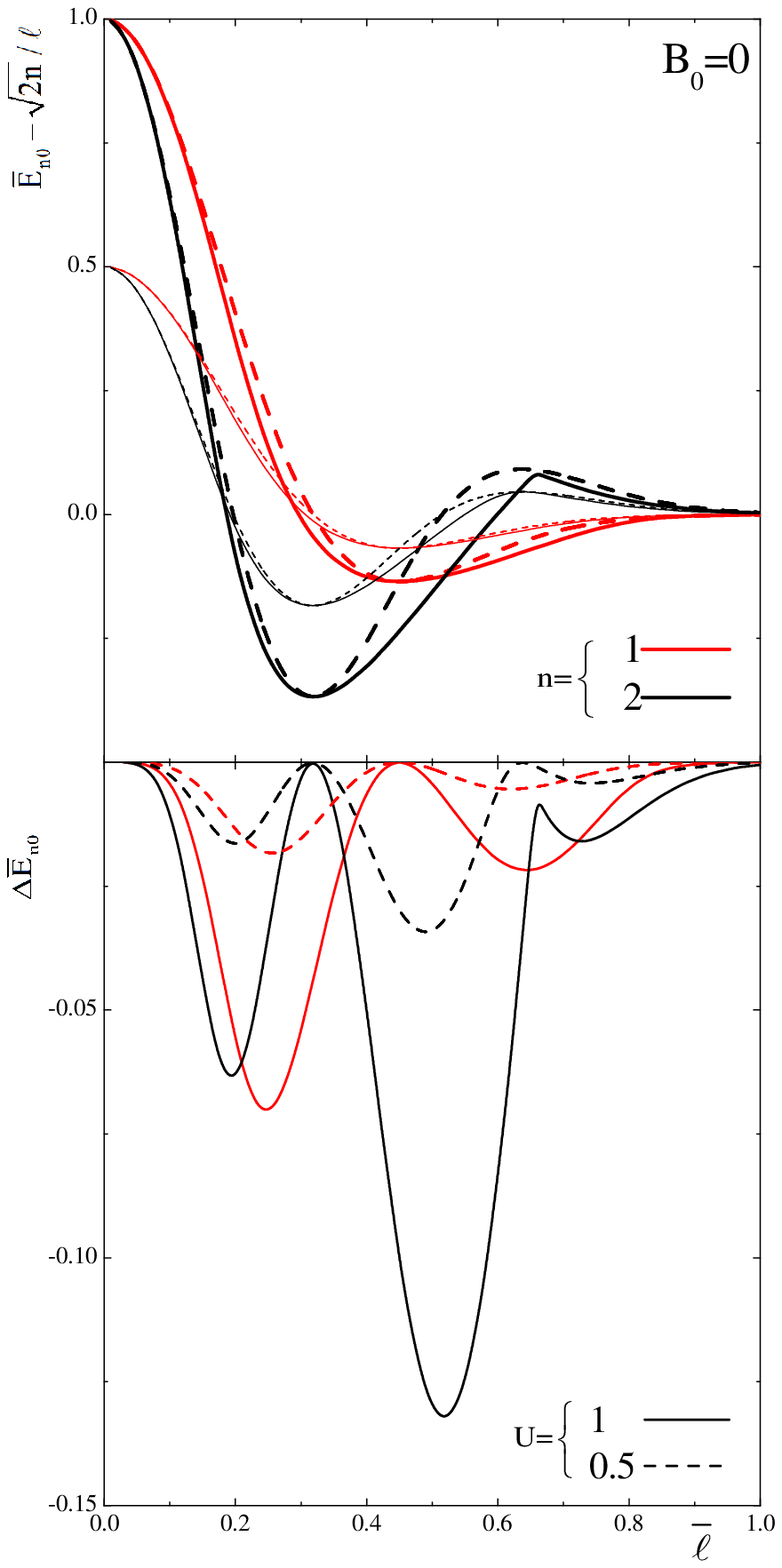} (a) %
\includegraphics*[  height=14cm,width=8cm]{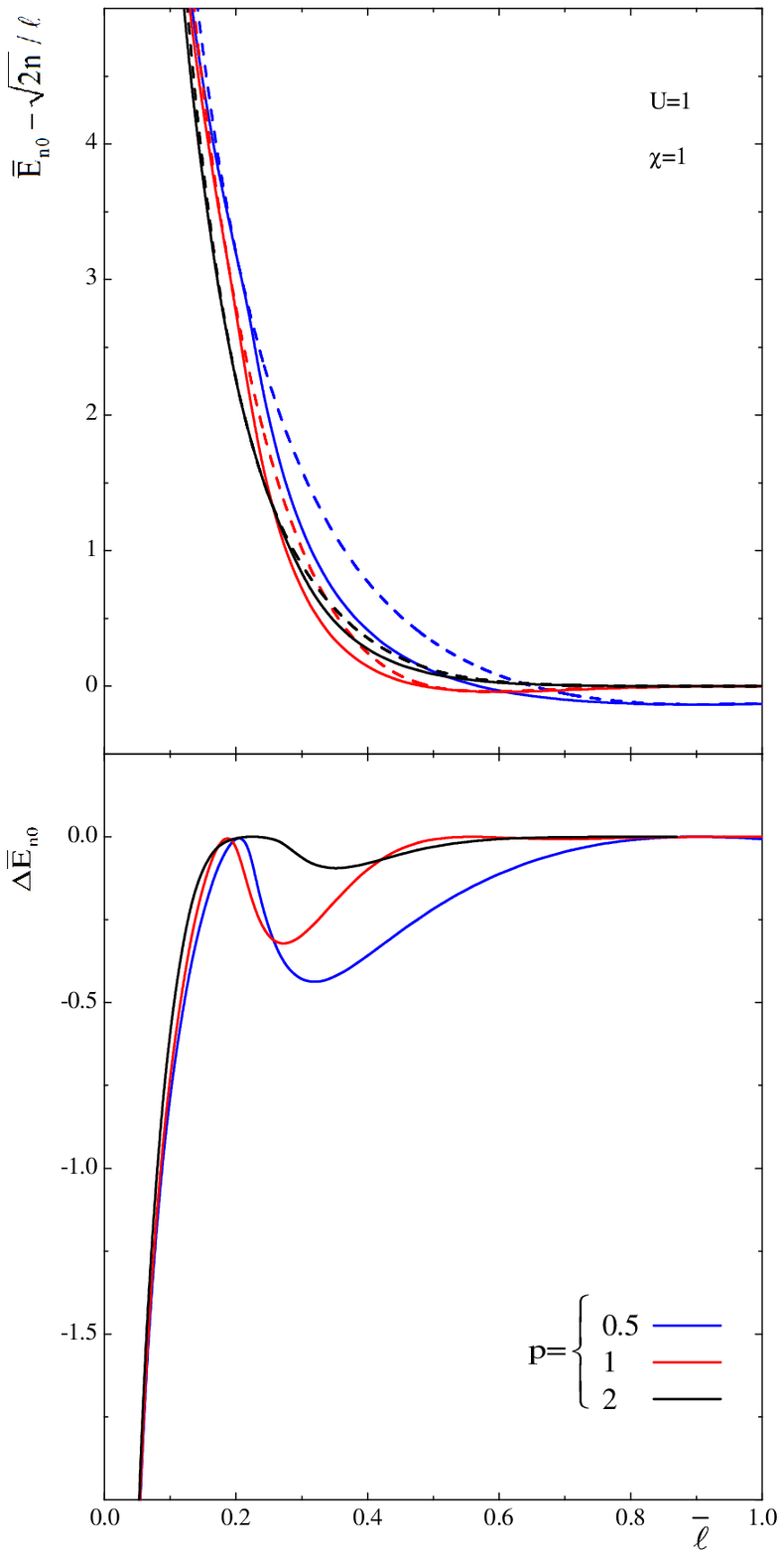} (b)
\caption{ (a) $\overline{E}_{no}-\protect\sqrt{2n}/\overline{\ell
}$ as a function of dimensionless magnetic confinement length
$\overline{\ell }=\ell /a_{0}$ for the first two Landau levels,
i.e. for $n=1,2$, in the absence of magnetic modulation for $U=1$
and $U=0.5$ with $\overline{k}_{y}=0$. While the solid curves
represent the present theoretical calculations, the dashed ones
corresponds calculations using first order non-degenerate
perturbation theory presented in Ref.~\onlinecite{peeters}.  In
upper panel, while upper curves  represent the $U=1$, lower curves
corresponds to the $U=0.5$. (b) The same as in (a), but for $U=1$
and $\protect\chi =1$. In both figures, the two theoretical curves
are compared with each other by plotting their
difference $\Delta \overline{E}_{n0}=\overline{E}_{n\overline{k}_{y}}^{V}-%
\overline{E}_{n\overline{k}_{y}}^{PT}$ as a function of dimensionless
magnetic confinement length $\overline{\ell }$ where $\overline{E}_{n%
\overline{k}_{y}}^{V}$ and $\overline{E}_{n\overline{k}_{y}}^{PT}$ denote
the results of present work and non-degenerate perturbation theory,
respectively (lower panels).}
\label{FIG1}
\end{figure}

The time independent effective massless Dirac-Weyl Hamiltonian we consider
is $H=H_{0}+U\left( \mathbf{r}_{\perp }\right) $, where
\begin{equation}
H_{0}=v_{F}\ \boldsymbol{\alpha }\cdot \left[ \mathbf{p}\left( \mathbf{r}%
\right) \mathbf{+}\frac{e}{c}\mathbf{A}\left( \mathbf{r}\right) \right] \ ,\
\boldsymbol{\alpha }\mathbf{=}%
\begin{pmatrix}
\boldsymbol{\sigma } & ~0 \\
0 & -\boldsymbol{\sigma }%
\end{pmatrix}
\label{1}
\end{equation}%
is the Hamiltonian for an electron which is minimally coupled to magnetic
field $\mathbf{B=}(0,0,B+B_{0}\cos Kx)$ with $K=2\pi /a_{0},$ through the
vector potential $\mathbf{A}$\textbf{=}$\mathbf{(}0\mathbf{,}%
Bx+(B_{0}/K)\sin Kx,0\mathbf{).}$ Here, $U\left( \mathbf{r}_{\perp }\right) $
is the one dimensional periodic electrostatic potential, and is given by $%
U\left( \mathbf{r}_{\perp }\right) =U\cos \left( K^\prime x\right) $ with $%
K^\prime =2\pi /c^\prime. $Here, $a_{0}$ and $c^\prime $ are the
periods of magnetic and electrostatic modulations, respectively.
In Eq.(\ref{1}), we have used the Dirac-Pauli representation of
Dirac matrix $\boldsymbol{\alpha }$ which is written of two by
two-block form in terms of Pauli spin matrices
$\boldsymbol{\sigma }$. Using two component spinor (pseudospin) as $%
\boldsymbol{\Psi }^{\dag }\boldsymbol{=}%
\begin{pmatrix}
\phi ^{\ast } & \chi ^{\ast }%
\end{pmatrix}%
$ we see that each component of eigenvalue equation $H_{0}\Psi =E\Psi $
satisfies the following coupled first order differential equations:

\begin{eqnarray}
\left( \boldsymbol{\sigma }\mathbf{\cdot p+}\frac{e}{c}\boldsymbol{\sigma }%
\mathbf{\cdot A}\right) \chi -\left[ \widetilde{E}+U\left( \mathbf{r}_{\perp
}\right) \right] \phi  &=&0  \notag \\
\left( \boldsymbol{\sigma }\mathbf{\cdot p+}\frac{e}{c}\boldsymbol{\sigma }%
\mathbf{\cdot A}\right) \phi -\left[ \widetilde{E}+U\left( \mathbf{r}_{\perp
}\right) \right] \chi  &=&0,  \label{2}
\end{eqnarray}%
where $\widetilde{E}=E/v_{F}$. Decoupling them in the absence of external
potentials, Eq.~(\ref{2}) yields the second-order differential equation
\begin{equation}
H_{0}^{^{\prime }}\phi \left( \mathbf{r}\right) =\left\{ \left[
p_{x}^{2}+\left( p_{y}+\frac{eB}{c}x\right) ^{2}\right] +\frac{e\hbar B}{c}%
\sigma _{3}-\widetilde{E}^{2}\right\} \phi \left( \mathbf{r}\right) ,
\label{3}
\end{equation}%
where $\boldsymbol{\sigma }_{3}$ is the third component of the Pauli
matrices. Since $\left[ H_{0}^{^{\prime }},p_{y}\right] =0$, we set
\begin{equation}
\phi \left( \mathbf{r}\right) =\frac{1}{\sqrt{L_{y}}}\exp \left(
ik_{y}y\right) \phi (x).  \label{4}
\end{equation}%
Therefore, Eq.~(\ref{3}) reduces to the solution of second order equation
for two-component wave function $\phi \left( x\right) $
\begin{equation}
\left\{ \frac{d^{2}}{dx^{2}}-\left( k_{y}+\frac{eB}{\hbar c}x\right) ^{2}+%
\left[ \left( \frac{\widetilde{E}}{\hbar }\right) ^{2}-\frac{eB}{\hbar c}%
\sigma _{3}\right] \right\} \phi \left( x\right) =0.  \label{5}
\end{equation}%
It is easy to show that, in pseudospin basis, the Hamiltonian
$H_{0}^{^{\prime }}$ in two-dimensions has solutions which can
easily be expressed in terms of the
Hermite polynomials by just setting with $\lambda =\left( \hbar c/eB\right) %
\left[ \left( \widetilde{E}/\hbar \right) ^{2}-\left( eBs/\hbar c\right) %
\right] =2\nu +1$. Therefore, the total solution of Eq.~(\ref{2}) can be
written as
\begin{equation}
\Psi _{+}=\frac{\exp \left( ik_{y}y\right) }{\sqrt{L_{y}}}%
\begin{bmatrix}
\mathbb{I}_{n-1}^{{}}\left( \gamma \right)  \\
0 \\
0 \\
i\mathbb{I}_{n}^{{}}\left( \gamma \right)
\end{bmatrix}%
,\Psi _{- }=\frac{\exp \left( ik_{y}y\right) }{\sqrt{L_{y}}}%
\begin{bmatrix}
0 \\
\mathbb{I}_{n}^{{}}\left( \gamma \right)  \\
-i\mathbb{I}_{n-1}^{{}}\left( \gamma \right)  \\
0%
\end{bmatrix}
\label{6}
\end{equation}%
and with
\begin{equation}
\mathbb{I}_{n}^{{}}\left( \gamma \right) =\left( \frac{\gamma ^{2}}{\pi }%
\right) ^{1/4}\frac{1}{\sqrt{2^{n}n!}}\exp \left\{ -\gamma ^{2}\left[
x+x_{0}^{{}}\right] ^{2}/2\right\} H_{n}\left[ \gamma \left(
x+x_{0}^{{}}\right) \right]   \label{8}
\end{equation}%
for the eigenvalues
\begin{equation}
\overline{E}_{n}^{\left( +\right) }=\sqrt{2neB/\hbar c}  \label{9}
\end{equation}%
which, in fact, corresponds to the eigenenergy  $\overline{E}_{n}^{\left(
+\right) }=\sqrt{2n}/\overline{\ell }$ for the  $\pi ^{\ast }$ band of
graphene, in which the quantum number $n$ takes on the values $n=\nu +\left(
s+1\right) /2$. One can follow the same procedure to obtain the associated
two-component solutions of Eq.~(\ref{1}) for the hole part of spectrum,
i.e., the $\pi $ band of graphene corresponding to the eigenenergy $%
\overline{E}_{n}^{\left( -\right) }=-\sqrt{2n}/\overline{\ell }$. In Eqs.~(%
\ref{5}-\ref{8}), we have defined  $x_{0}^{{}}=c\hbar k_{y}/eB$, and
rescaled the energy $\widetilde{E}$ by dividing it by $\hbar $ to obtain $%
\overline{E}=\widetilde{E}$ /$\hbar =E/\hbar v_{F}$, and defined
$\gamma ^{2}=eB/\hbar c$, $\overline{\ell }^{2}=\hbar c/eB$. Then,
for the value $s=1 $  we obtain $\nu =n-1$, and for \ $s=-1$ we
obtain $\nu =n$. It should be noted that, $s=\pm 1$ does not
represent "spin up" and "spin down", but they describe states on
the A(B) sublattice of graphene. In the presence of external
potentials, to calculate their combined effects onto eigenvalues
given by Eq.~(\ref{9}), we choose Eq.~(\ref{6}) and Eq.~(\ref{8})
as basis functions for our variational procedure. Therefore, in
this sense one has to minimize the energy,
\begin{equation}
\overline{E}^{+}\left( \gamma \right) =\int d^{2}\mathbf{r}\Psi _{\pm
}^{\dag }\left( \gamma \right) \left[ \boldsymbol{\alpha }\mathbf{\cdot p+}%
\frac{e}{\hbar c}\boldsymbol{\alpha }\mathbf{\cdot }\left( 0,Bx+\frac{B_{0}}{%
K}\sin Kx,0\right) +\overline{U}\left( \mathbf{r}_{\perp }\right) \right]
\Psi _{\pm }\left( \gamma \right) ,  \label{10}
\end{equation}%
where $\overline{U}\left( \mathbf{r}_{\perp }\right) =U\left( \mathbf{r}%
_{\perp }\right) /\hbar v_{F}$.
Analogously, one can follow the
same treatment for the negative energy
pseudospinors  to obtain $%
\overline{E}^{-}\left( \gamma \right) $.\ Finally, performing the
related integrals in Eq.~(\ref{10}), we find that
\begin{eqnarray}
\overline{E}_{n\overline{k}_{y}}^{\pm }\left( \overline{\gamma }\right)  &=&%
\frac{\sqrt{2n}}{2}\overline{\gamma }+\frac{\sqrt{2n}}{2\cdot \overset{-}{%
\ell }^{2}\overline{\gamma }}+\frac{\sqrt{2n}}{2\pi ^{2}\cdot \overline{\ell
}_{0}^{2}}\overline{\gamma }\exp \left( -\pi ^{2}/\overline{\gamma }%
^{2}\right) \cos \left( 2\pi \overline{\ell }^{2}\overline{k}_{y}\right)
\notag \\
&&\times \left[ L_{n-1}\left( \frac{2\pi ^{2}}{\overline{\gamma }^{2}}%
\right) -L_{n}\left( \frac{2\pi ^{2}}{\overline{\gamma }^{2}}\right) \right]
+\frac{\overline{U}}{2}\cdot \cos \left( 2\pi p\overline{\ell }^{2}\overline{%
k}_{y}\right)   \notag \\
&&\times \left[ L_{n}\left( p^{2}\frac{2\pi ^{2}}{\overline{\gamma }^{2}}%
\right) +L_{n-1}\left( p^{2}\frac{2\pi ^{2}}{\overline{\gamma }^{2}}\right) %
\right] \exp \left( -p^{2}\pi ^{2}/\overline{\gamma }^{2}\right)   \label{12}
\end{eqnarray}%
where the energy $\overline{E}_{n\overline{k}_{y}}^{\pm }\left( \overline{%
\gamma }\right) $ is again rewritten, for convenience, in the units of $%
\hbar v_{F}$, and all the lengths are in the units of $a_{0}$, $\overline{%
\gamma }=\gamma a_{0}$, $\overline{\ell }=\ell /a_{0}$, $\overline{\ell} _{0}%
=\ell _{0}/a_{0}$, $\overline{k}_{y}=k_{y}a_{0}$. Here, we have
also defined $p=a_{0}/c^{\prime }$. Therefore, our rescaled
variables in Eq.~(\ref{12}) measure lengths in the units of
$a_{0}$ and energies in the units of $\hbar v_{F}$.
Eq.~(\ref{12}), when minimized with respect to $\overline{\gamma
}$, gives the effect of external one dimensional electrostatic
potential together with its magnetic analogue onto the Dirac-Weyl
like Landau levels given by Eq.~(\ref{9}). To show this, we start
from the well-known case, absence of external fields, i.e.,
$B_{0}$, $U.$ In this case, minimization
of Eq.~(\ref{12}) with respect to $\overline{\gamma }$ yields $\overline{%
\gamma }^{2}=eB/\hbar c$ $=1/\overline{\ell }^{2}$. Replacing this result
back into Eq.~(\ref{12}) yields
\begin{eqnarray}
\overline{E}_{n\overline{k}_{y}}\left( \overline{\gamma }\right)  &=&\frac{%
\sqrt{2n}}{\overline{\ell }}+\frac{1}{2\pi ^{2}}\sqrt{\frac{n}{2}}\frac{1}{%
\overline{\ell }_{0}^{2}\overline{\ell }}\exp \left( -\pi ^{2}\overline{\ell
}^{2}\right) \cos \left( 2\pi \overline{\ell }^{2}\overline{k}_{y}\right)
\notag \\
&&\times \left[ L_{n-1}\left( 2\pi ^{2}\overline{\ell }^{2}\right)
-L_{n}\left( 2\pi ^{2}\overline{\ell }^{2}\right) \right] +\frac{\overline{U}%
}{2}\cos \left( 2\pi p\overline{\ell }^{2}\overline{k}_{y}\right)   \notag \\
&&\times \left[ L_{n}\left( 2\pi ^{2}p^{2}\overline{\ell }^{2}\right)
+L_{n-1}\left( 2\pi ^{2}p^{2}\overline{\ell }^{2}\right) \right] \exp \left(
-p^{2}\pi ^{2}\overline{\ell }^{2}\right) .  \label{14}
\end{eqnarray}%
It is easy to show that Eq.~(\ref{14})  reduces exactly to those
found in Refs.~[\onlinecite{peeters,tahir}] by using first-order
perturbation correction. Of course, it covers inherently
well-known Dirac like Landau levels,
$\overline{E}^{+}=+\sqrt{2n}/\overline{\ell }$ in the absence of
external fields. In FIG.~\ref{FIG1}(a), we plot dimensionless
magnetic confinement length $\overline{\ell }$ variation of the
dimensionless half bandwidth
$\overline{E}_{n0}-\sqrt{2n}/\overline{\ell }$ for two different
values of $U=0.5$ and $1$ resulting from the numerical minimization of Eq.~(%
\ref{12}) (solid lines), together with the results of non-degenerate
first-order perturbation calculation (dashed lines) as well as their
difference $\Delta \overline{E}$ (lower panel). In order to observe the
effect of one dimensional magnetic analogue in FIG.~\ref{FIG1}(b) we also
plot the same in FIG.~\ref{FIG1}(a) but for $\chi =1$ ($\chi =B_{0}/U$) with $%
p=0.5$, $1$, and $2$, respectively. From both figures, as we decrease $%
\overline{\ell }$ ,i.e, increase $\overline{B}=B/\widetilde{B}$ ($\widetilde{%
B}=\hbar c/ea_{0}^{2}$) Landau levels broaden into mini subbands whose
boundaries are determined by two asymptotic values of $\overline{k}_{y}$,
i.e, by $\overline{k}_{y}=0$ and $\overline{k}_{y}=1/2\pi \overline{\ell }.$
Of these only with $n=1$ and $n=2$ are depicted in FIG.~\ref{FIG1}(a) for $%
\overline{k}_{y}=0$.

It is clear from FIG.~\ref{FIG1}(a) and (b) that, as $\overline{\ell }$
decreases, dimensionless half bandwidth of the first two Landau levels
begins to highly oscillate, due to the oscillatory nature of Laguerre
polynomials, passing through degeneracy restoring points, i.e., wherein the
flatband condition is fulfilled, and thus Weiss oscillations are suppressed.
It should also be noted that, from the lower panels, a rapid increase of $%
\Delta \overline{E}$ with increase in $U$ is evident, yielding significant
discrepancies above $0.1\hbar v_{F}.$ Moreover, note that $\Delta \overline{E%
}$ diminishes at some values of $\overline{\ell }$ where
approximately flatband condition is fulfilled. In other words,
since diminishing $\Delta \overline{E}$ indicates that the
variational and the perturbational results are almost same, we can
say that, around these points, i.e., where the flatband condition
is fulfilled, perturbational results can be safely used, but
otherwise they cannot.

\begin{figure}[p]
\includegraphics*[ height=10cm,width=10cm]{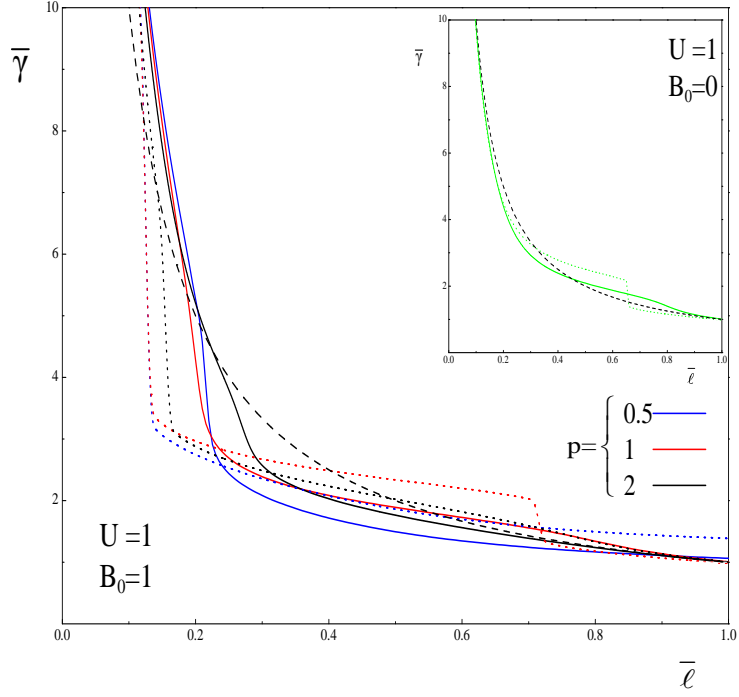} %
\caption{The dimensionless variational parameter $\overline{\gamma
}$ as a function of dimensionless magnetic confinement length
$\overline{\ell}$. While the solid lines correspond to energy
level with $n=1$, the dotted lines correspond to energy level with
$n=2$, for various sets of $p$. The inset shows the $B_0=0$ case
for $U=1$.} \label{FIG2}
\end{figure}

We further plot in FIG~\ref{FIG2} the variations of dimensionless
variational parameter $\overline{\gamma}$ as a function of
dimensionless magnetic confinement length $\overline{\ell}$ for
the first two Landau levels with different $p$ values. Also, for
comparison, the curve for the unperturbed case, i.e.,
$\overline{\gamma }=1/\overline{\ell }$, is plotted (dashed bold
line). We note two features of these curves. First,  there exist
discrepancies between the unperturbed and perturbed curves in the
high magnetic field regime, $\overline{\ell }<1$ where the
variational picture becomes more appropriate. Second, again in
this region the perturbed curves intersect the unperturbed one
wherein the flatband condition is fulfilled, as is indicated
above. These justify the validity of our variational procedure.

\begin{figure}[p]
\includegraphics*[ height=14cm,width=8cm]{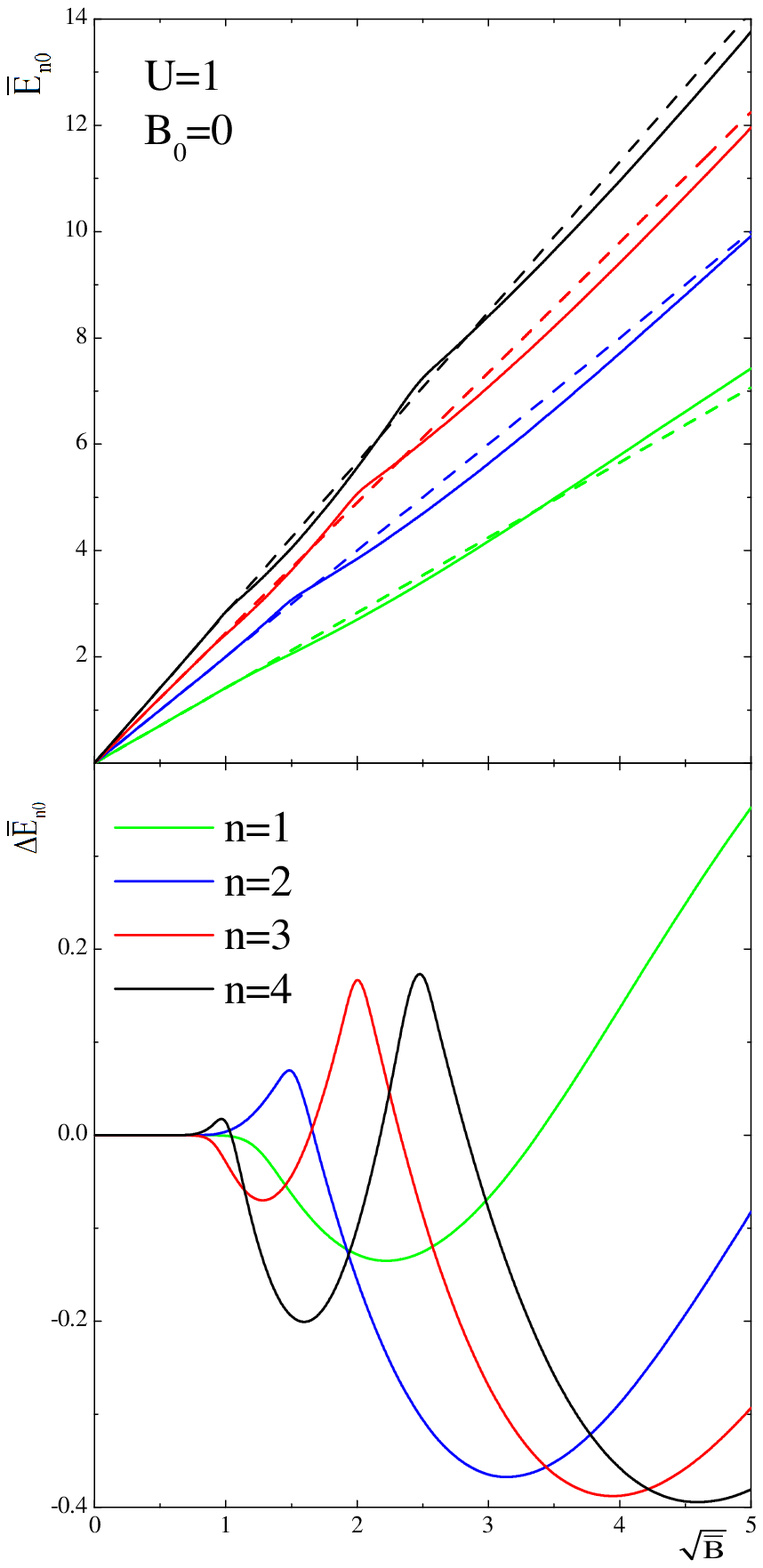}(a) %
\includegraphics*[ height=14cm,width=8cm]{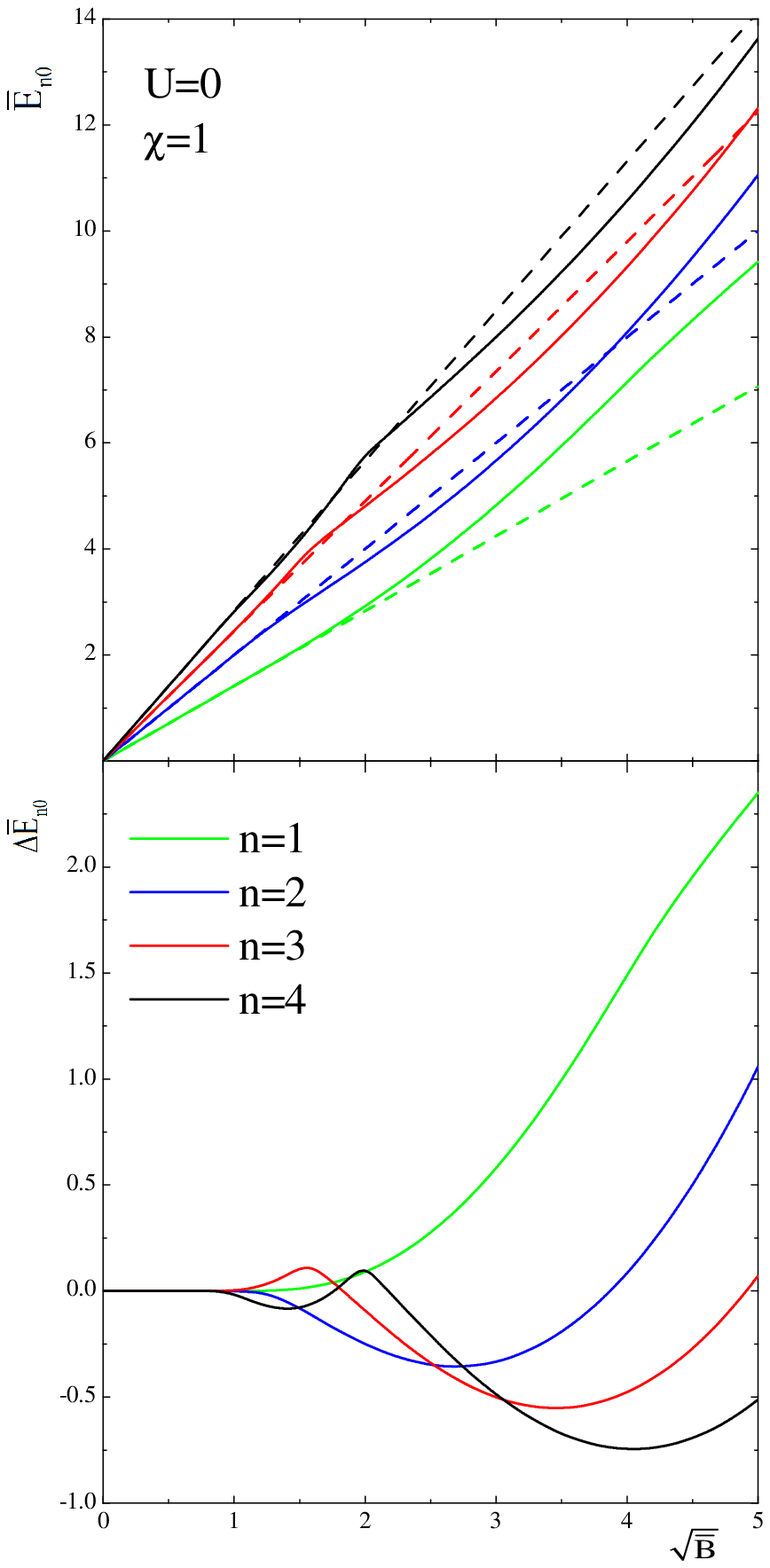}(b)
\caption{(a) The first four low-lying Landau levels as a function of $%
\protect\sqrt{\overline{B}}$ with $U=1$ for $B_{0}=0$, and for $\overline{k}%
_{y}=0$. While the dashed curves correspond to unperturbed Landau levels,
i.e.,$\protect\sqrt{2n\overline{B}}$, the solid ones denotes the effect of
electrostatic potential, (upper panel). To demonstrate the effect of
external modulated electrostatic potential $\overline{E}_{n0}-\protect\sqrt{%
2n\overline{B}}$ is also plotted as a function of $\protect\sqrt{\overline{B}%
}$, in lower Panel. (b) The same as in (a), but for $U=0$ and $\protect\chi %
=1$. } \label{FIG3}
\end{figure}

In FIG.~\ref{FIG3}, $\overline{E}_{n0}$ as a function of
$\sqrt{\overline{B}}$ are plotted for the first four Landau
levels. With increasing $\overline{B}$ strong deviations from the
unperturbed Dirac-Weyl like Landau levels (dashed ones) appear
near $\overline{B}\sim 1$. To see these effects more clearly,
variations of half-bandwidths of the corresponding levels, i.e.,
$\Delta
\overline{E}_{n0}=\overline{E}_{n0}-\sqrt{2n\overline{B}}$, are
given as a function of $\sqrt{\overline{B}}$ in the lower panel of
the same figure.

\begin{figure}[p]
\includegraphics* [height=12cm,width=8cm]{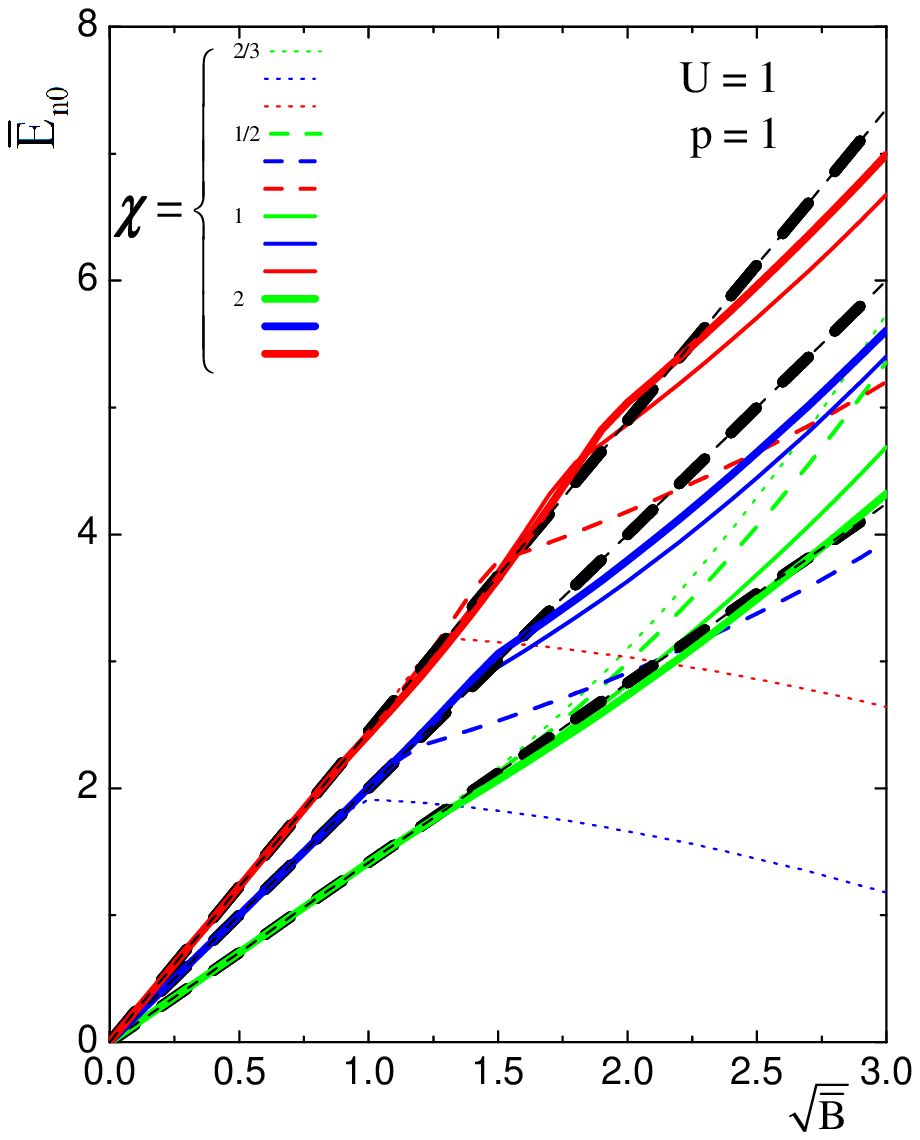} (a) %
\includegraphics* [height=12cm,width=8cm]{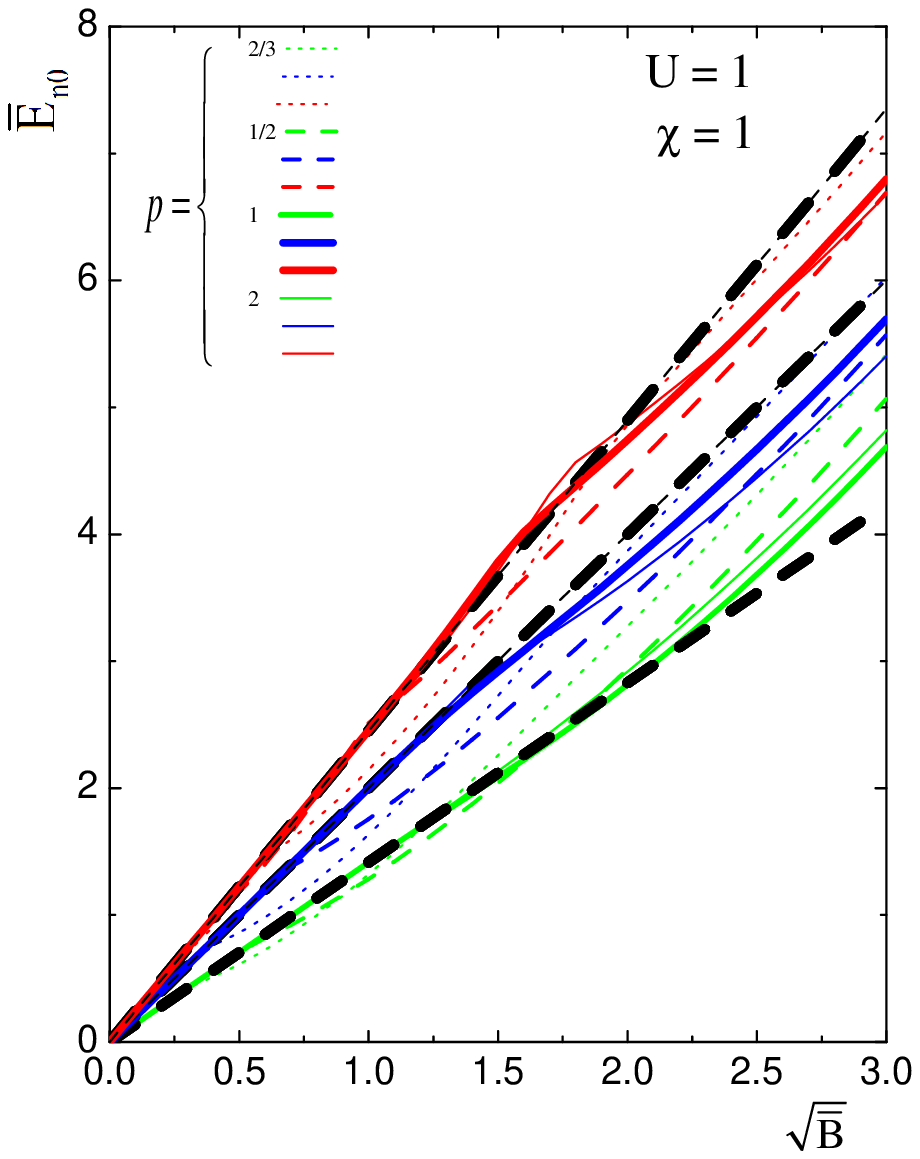} (b)
\caption{The same as in FIG.~\protect\ref{FIG2}, but for various
sets of $p$ and $\protect\chi $, and only for the first three
energy levels. The bold dashed straight lines refer to the
relevant unperturbed Landau levels} \label{FIG4}
\end{figure}

Finally, in FIG.~\ref{FIG4}(a) and (b) we have plotted
$\overline{E}_{n0}$ as a function of $\sqrt{\overline{B}}$ for the
four $p$ values, and for the
four $\chi $ values, respectively. It should be noted that the variable $%
p=a_{0}/c^{\prime }$ measure whether oscillating fields are in phase or out
of phase, and the variable $\chi $ measures the ratio of amplitudes of
fields. One observes very clearly that, by changing their phases and/or
their ratios of amplitudes, dramatic changes in the spectrum occur when $%
\overline{B}$ decreases.

In conclusion, since our variational method yields energies lower
than that obtained by the first-order perturbation calculation, it
provides an efficient procedure for finding out the effects of
external potentials onto the electronic spectrum of a graphene
electron in an uniform magnetic field. Moreover, we see that,
after performing various analytical checks on the obtained energy
spectra, our results cover the well-known results found in the
literature \cite{peeters,tahir},  and in the absence of external
fields they reduce to the $2D$-Landau levels of massless Dirac
fermions in some certain values of variational parameter. We use
the solutions of massless Dirac fermions in an uniform magnetic
field as trial wave functions to obtain the low-lying energy
spectrum of these particles in the presence of external
potentials. For Landau bands with high indices in the case of high
magnetic fields, adjacent Landau bands may overlap. In this case,
the variational function should be a linear combination of these
sates rather than one of them. In this regime, as a consequence of
overlapping,  even-odd transitions in the Shubnikov-de Haas
oscillations\cite{Shi} may be expected in the graphene.

\begin{acknowledgments}
The authors thank Professor T. Altanhan for valuable discussions, and for a
critically reading of the manuscript.
\end{acknowledgments}

\end{document}